\documentclass[aps,amsmath,amssymb,preprint]{revtex4}
\usepackage{graphicx}% Include figure files
\usepackage{dcolumn}% Align table columns on decimal point
\usepackage{bm}% bold math

\begin{document}
\title{Superconductivity at 43 K in Samarium-arsenide Oxides
$SmFeAsO_{1-x}F_x$ }
\author{X. H. Chen}
\altaffiliation{Corresponding author} \email{chenxh@ustc.edu.cn}
\author{T. Wu, G. Wu, R. H. Liu, H. Chen and D. F. Fang}
\affiliation{Hefei National Laboratory for Physical Science at
Microscale and Department of Physics, University of Science and
Technology of China, Hefei, Anhui 230026, People's Republic of
China\\}

\date{\today}
\maketitle

{\bf  Since the discovery of high-transition temperature ($T_c$)
superconductivity in layered copper oxides, extensive efforts have
been devoted to explore the higher $T_c$ superconductivity. However,
the $T_c$ higher than 40 K can be obtained only in the copper oxide
superconductors so far. The highest reported value of $T_c$ for
non-copper-oxide bulk superconductivity is 39 K in
$MgB_2$.\cite{jun} The $T_c$ of about 40 K is close to or above the
theoretical value predicted from BCS theory.\cite{mcmillan}
Therefore, it is very significant to search for non-copper oxide
superconductor with the transition temperature higher than 40 K to
understand the mechanism of high-$T_c$ superconductivity. Here we
report the discovery of bulk superconductivity in samarium-arsenide
oxides $SmFeAsO_{1-x}F_x$ with ZrCuSiAs type structure. Resistivity
and magnetization measurements show strong evidences for transition
temperature as high as 43 K. $SmFeAsO_{1-x}F_x$ is the first
non-copper oxide superconductor with $T_c$ higher than 40 K. The
$T_c$ higher than 40 K may be a strong argument to consider
$SmFeAsO_{1-x}F_x$ as an unconventional superconductor. }

Layered rare-earth metal oxypnictides LaOMPn (M=Fe, Co, Ni, Ru and
Pn=P and As) have attracted great attention recently due to the
discovery of superconductivity at $T_c=26$ K in the iron-based
$LaO_{1-x}F_x$FeAs (x=0.05-0.12).\cite{yoichi} $LaO_{1-x}F_x$FeAs
adopts ZrCuSiAs type structure. The quaternary equiatomic ZrCuSiAs
type structure is very simple with only eight atoms in the
tetragonal cell. A series of equiatomic quaternary compounds RFeAsO
and RFePO (R=La-Nd, Sm, Gd) with  ZrCuSiAs type structure has been
reported by Quebe et al. and Zimmer et al.\cite{quebe,zimmer} The
crystal structure of the tetragonal ZrCuSiAs type compound
SmFeAs(O,F) is shown in Fig.1.

The polycrystalline samples with nominal composition
$SmFeAsO_{1-x}F_{x}$ (x=0.15) were synthesized by conventional solid
state reaction using high purity SmAs, $SmF_3$, Fe and $Fe_2O_3$ as
starting materials. SmAs was obtained by reacting Sm chips and As
pieces at 600 $^oC$ for 3 hours and then 900 $^oC$ for 5 hours. The
raw materials were thoroughly grounded and pressed into pellets. The
pellets were wrapped into Ta foil and sealed in an evacuated quartz
tube. They are then annealed at 1160 and 1200 $^oC$ for 40 hours,
respectively. The sample preparation process except for annealing
was carried out in glove box in which high pure argon atmosphere is
filled. The samples were characterized by X-Ray diffraction (XRD)
using Rigaku D/max-A X-Ray diffractometer with Cu K$\alpha$
radiation ($\lambda$=1.5418\AA) in the 2$\theta$ range of
10$-$70$^{\circ}$ with the step of 0.02$^{\circ}$ at room
temperature.

Figure 2 shows the XRD pattern for the sample annealed at 1160
$^oC$, respectively. It is found that the peaks in XRD pattern can
be well indexed to the tetragonal ZrCuSiAs-type structure with
a=0.3943 nm and c=0.8514 nm except for some tiny peaks from impurity
phase $SmOF$ for the sample with x=0.15 annealed at 1160 $^oC$, the
lattice parameters are slightly larger than a=0.3940 nm and c=0.8496
nm of F-free SmFeAsO due to substitution of large anion $F^-$ for
small anion $O^{2-}$.

The susceptibility measurement was performed in Quantum Design
MPMS-7T system (Quantum Design). The magnetic characterization of
the superconducting transitions is shown in Fig.3 under magnetic
field of 10 Oe in the zero-field cooled and field-cooled
measurements for the sample annealed at 1160 $^oC$. The magnetic
onset for the superconducting transition is 41.8 K for the sample
annealed at 1160 $^oC$, and 41.3 K for the sample annealed at 1200
$^oC$ (not shown), respectively. The existence of the
superconducting phase was unambiguously confirmed by the Meissner
effect on cooling in a magnetic field. A superconducting volume
fraction of about 50\% under a magnetic field of 10 Oe was obtained
at 5 K, indicating that the superconductivity is bulk in nature.

Resistivity measurements were performed on a AC resistance bridge
(Linear Research Inc., Model LR700) by the standard four-probe
method. Figure 4 shows the temperature dependence of the resistivity
under magnetic field H=0, 5 and 7 Tesla for the samples annealed at
1160 $^oC$ and 1200 $^oC$, respectively. Under zero field, the onset
transition and midpoint temperatures of the resistive transition are
43 K and 41.7 K for the sample annealed at 1160 $^oC$, 43.7 K and
41.2 K for the sample annealed at 1200 $^oC$, respectively. The
90\%-10\% transition width is 2.5 K and 3 K for the samples annealed
at 1160 $^oC$ and 1200 $^oC$, respectively. It is found that the
onset transition temperature in susceptibility is consistent with
the transition midpoint temperature in resistivity. External
magnetic field of 5 and 7 Tesla makes the transition width
broadening, but the onset transition temperature is not sensitive to
magnetic field, indicating that the upper critical field is very
high for this superconductor. Therefore, this superconductor has
potential application due to high transition temperature and high
upper critical field.  Replacement of La by Sm leads to a big
increase in $T_c$ from 26 K in $LaO_{1-x}F_xFeAs$ \cite{yoichi} to
43 K in $SmFeAsO_{1-x}F_x$.  It suggests that it is possible to
realize higher $T_c$ in such layered oxypnictides. The $T_c$ of 43 K
in $SmFeAsO_{1-x}F_x$ higher than the theoretical value predicted
from BCS theory\cite{mcmillan} provides strong argument to consider
layered oxypnictide superconductors as an unconventional
superconductor.

\vspace*{2mm} {\bf Acknowledgment:} This work is supported by the
Nature Science Foundation of China and by the Ministry of Science
and Technology of China (973 project No: 2006CB601001) and by
National Basic Research Program of China (2006CB922005).

\newpage
\begin{figure}
\includegraphics[width=\textwidth]{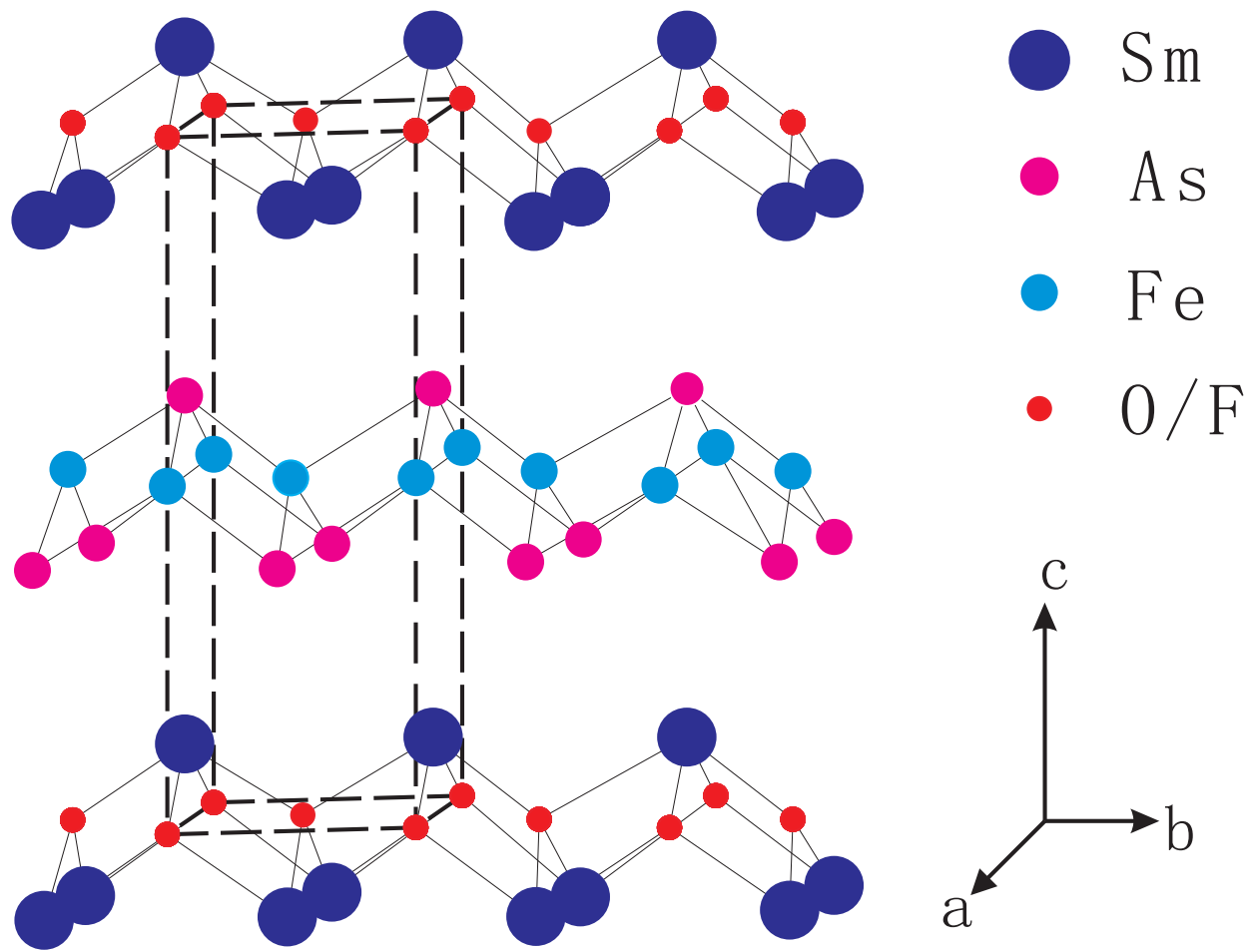}
\caption{Structural model of $SmFeAsO_{1-x}F_x$ with the tetragonal
ZrCuSiAs type structure.\\}
\end{figure}

\newpage
\begin{figure}
\includegraphics[width=\textwidth]{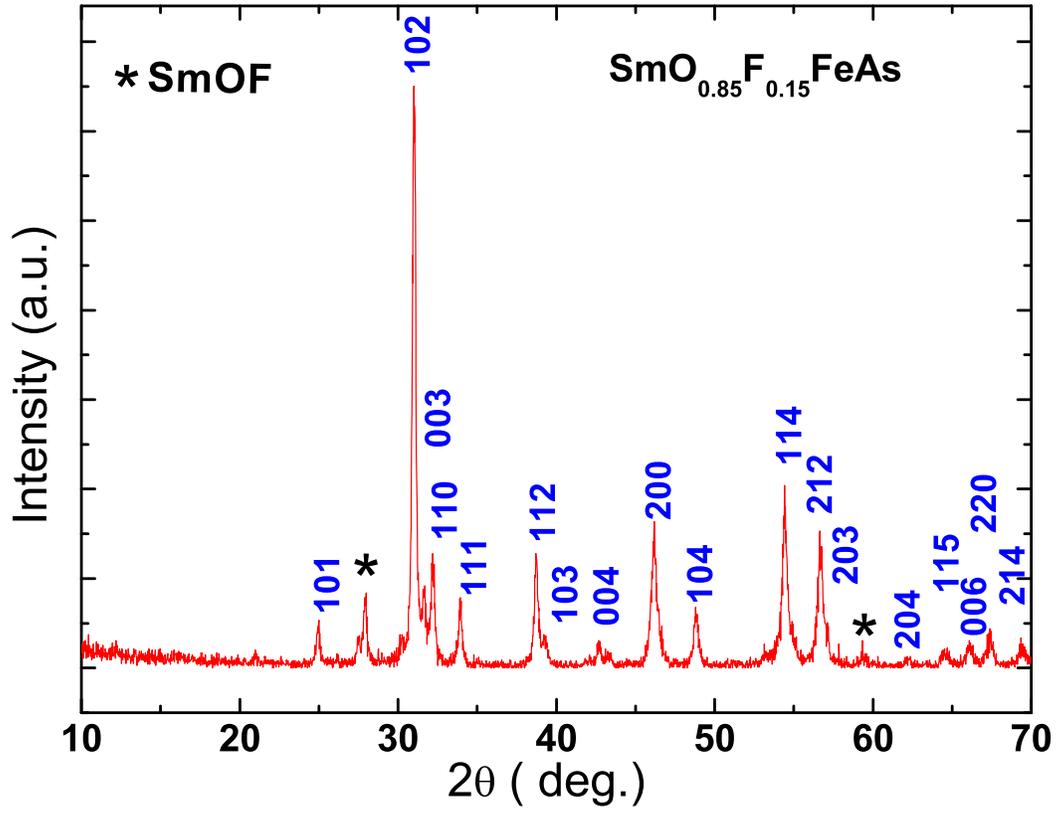}
\caption{ X-ray diffraction pattern for the sample with nominal
composition $SmFeAsO_{1-x}F_x$ (x=0.15) annealed at 1160 $^oC$. A
tiny impurity phase SmOF is observed.
\\}
\end{figure}

\newpage
\begin{figure}
\includegraphics[width=\textwidth]{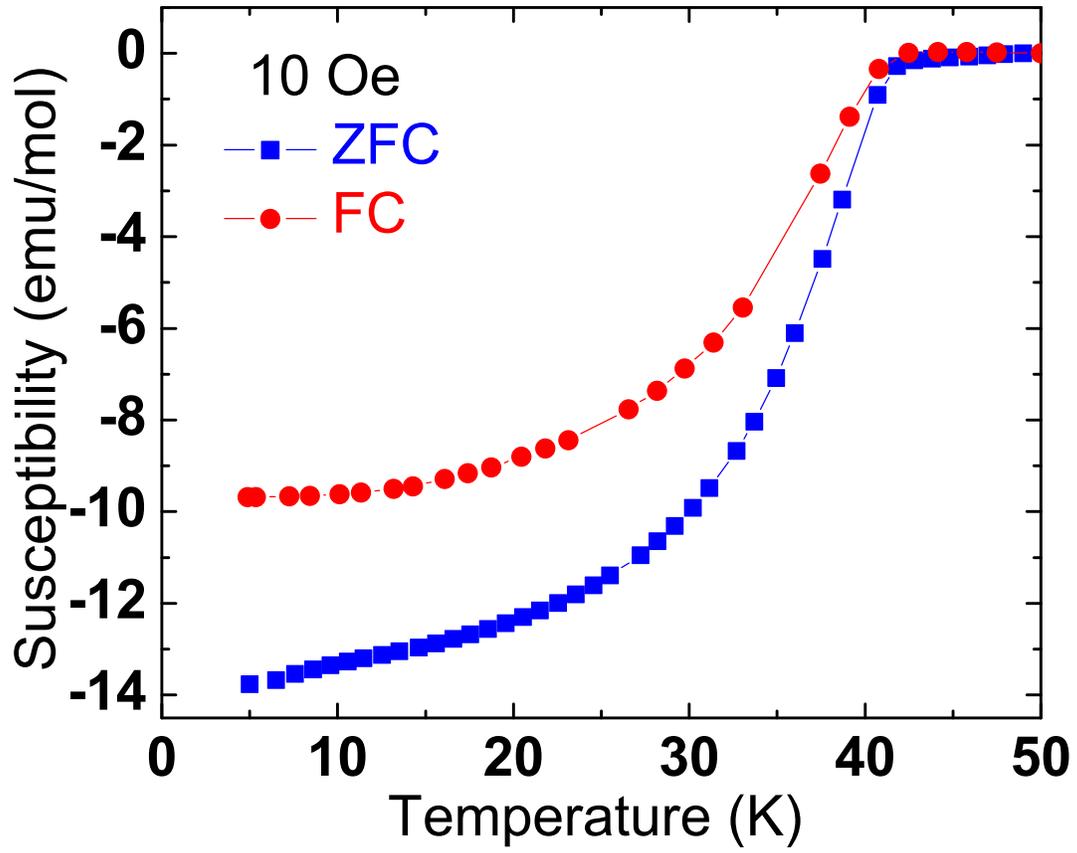}
\caption{Temperature dependence of magnetic susceptibility as a
function of temperature for the sample annealed at 1160 $^oC$ in the
zero-field cooled (ZFC) and field-cooled measurements at 10 Oe.
\\}
\end{figure}

\newpage
\begin{figure}
\includegraphics[width=0.8\textwidth]{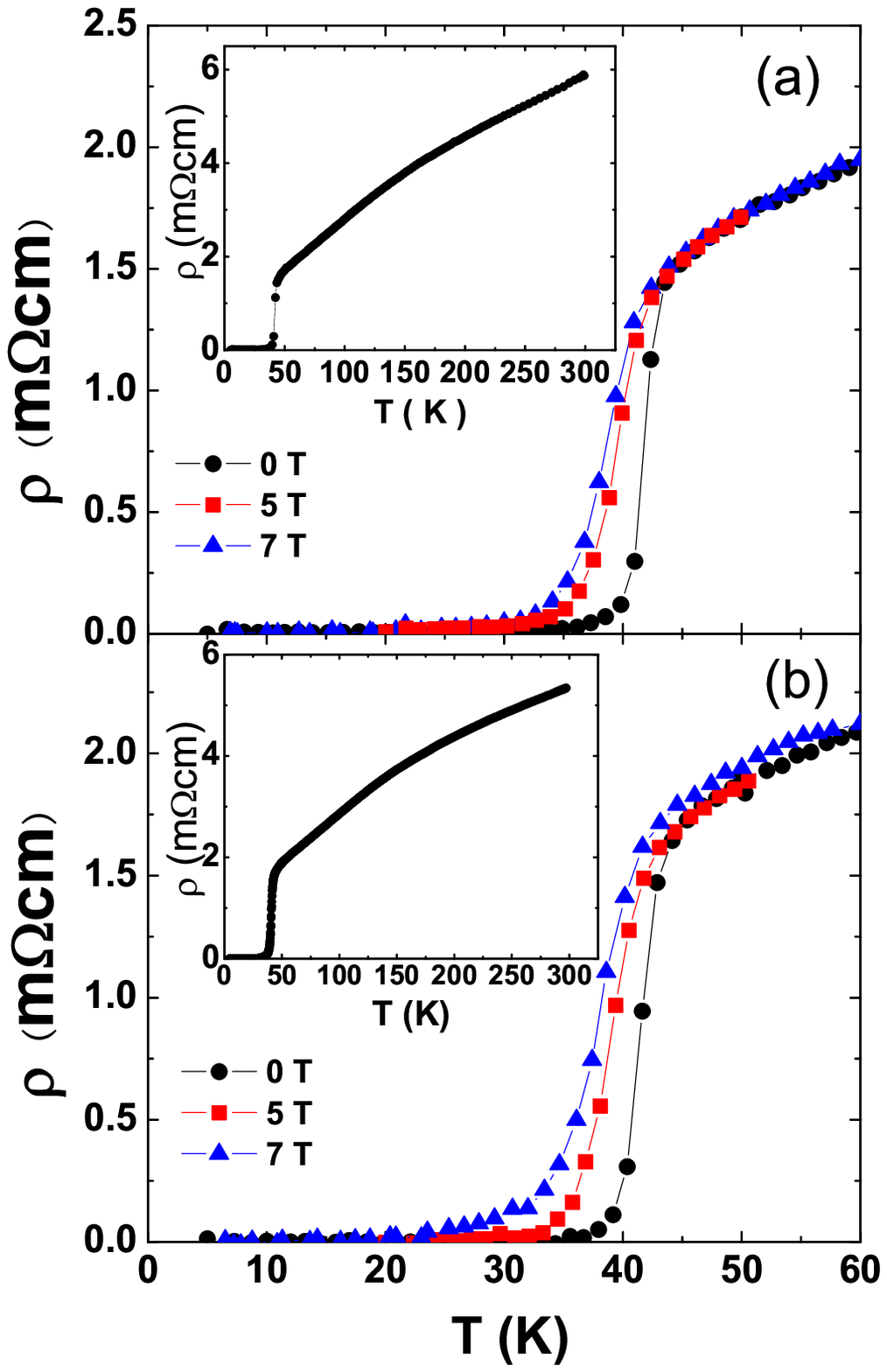}
\caption{Temperature dependence of resistivity without magnetic
field and under magnetic field of 5 and 7 T for the samples annealed
at 1160 $^oC$ (a) and 1200 $^oC$ (b), respectively. Inset shows the
resistivity from 300 K to 5 K.\\
}
\end{figure}

\clearpage
 \setcounter{figure}{0}

\renewcommand{\thefigure}{\arabic{figure}s}


\begin{thebibliography}{00}

\baselineskip 9pt

\bibitem{jun}
Nagamatsu, J. et al., Superconductivity at 39 K in magnesium
diboride. \emph{Nature} {\bf 410}, 63-64(2001).
\bibitem{mcmillan}
McMillan, W.L., Transition Temperature of Strong-coupled
superconductors. \emph{Phys. Rev.} {\bf 167}, 331-344(1968).
\bibitem{yoichi}
Kamihara, Y. et al., Iron-based layered superconductor
$LaO_{1-x}F_x$FeAs (x=0.05-0.12) with $T_c$=26 K. \emph{J. Am. Chem.
Sco.} {\bf 130}, 3296(2008).
\bibitem{quebe}
Quebe, P. et al., Quaternary rare earth transition metal arsenide
oxides RTAsO (T=Fe, Ru, Co) with ZrCuSiAs type structure. \emph{J.
Alloys and Compounds} {\bf 302}, 70-74(2000).
\bibitem{zimmer}
Zimmer, B. I. et al., The rare earth transition metal phosphide
oxides LnFePO, LnRuPO and  LnCoPO with ZrCuSiAs type structure.
\emph{J. Alloys and Compounds} {\bf 229}, 238-242(1995).


\end{thebibliography}
\end{document}